\documentclass{aa}
\begin{document}
\def\lae{\mathrel{<\kern-1.0em\lower0.9ex\hbox{$\sim$}}}
\def\gae{\mathrel{>\kern-1.0em\lower0.9ex\hbox{$\sim$}}}
\title{Optical identifications of High Frequency Peakers\thanks{Based
 on observations made with the Italian Telescopio Nazionale Galileo
 (TNG) operated on the island of La Palma by the Centro Galileo
 Galilei of the CNAA (Consorzio Nazionale per l'Astronomia e
 l'Astrofisica) at the Spanish Observatorio del Roque de los
 Muchachos of the Instituto de Astrofisica de Canarias}
} 

\author{D. Dallacasa\inst{1,2}
 \and R. Falomo\inst{3}
 \and C. Stanghellini\inst{4}}

\offprints{D. Dallacasa, \\  e-mail: ddallaca@ira.bo.cnr.it}

\institute{Dipartimento di Astronomia, Universit\`a degli Studi, Via Ranzani 1
          I-40127 Bologna, Italy
 \and Istituto di Radioastronomia - CNR, Via Gobetti 101, I-40129
      Bologna, Italy
 \and Osservatorio Astronomico, Vicolo dell'Osservatorio 5, I-35122
      Padova, Italy 
 \and Istituto di Radioastronomia - CNR, C.P. 141, I-96017 Noto (SR),
      Italy }

\date{Received 21 September 2001; Accepted 8 November 2001}

\abstract{
We present CCD observations of 13 objects from a complete sample of 55
bright High Frequency Peaker (HFP) radio sources, and provide optical 
identification for 12 of them.\\
Images in R and V filters have been used to derive some additional
information concerning the host of the radio source. Three hosts are
likely to be galaxies, one resulted slightly extended, while the
remaining 8 are likely distant quasars. Based on these identifications
and those available in the literature, we find that the fraction of
quasars in our HFP sample is significantly higher than in samples of
Compact Steep-Spectrum and GHz-Peaked Spectrum radio sources.
}

\maketitle

\section{Introduction}

The population of {\bf intrinsically small} and {\bf bright} radio
sources has been investigated in the past by studying samples of
Compact Steep Spectrum (CSS) and GHz-Peaked Spectrum (GPS) radio
sources (see Spencer et al. 1989, Fanti et al. 1990, 1995,
Stanghellini et al. 1998, O'Dea 1998). These sources are characterized
by optically thin steep radio emission with a turnover occurring at
frequencies ranging from about 100 MHz (CSSs) to about 1 GHz
(GPSs). When the spectral peak is found at frequencies higher than a
few GHz, the source is classified High Frequency Peaker (HFP,
Dallacasa el al. 2000).\\
There is a substantial absorption of the low frequency radiation,
but it is unclear whether it is due either to synchrotron
self-absorption (SSA) favoured by  Fanti et al. (1995) and Snellen et
al. (2000) or to free-free absorption (FFA) as proposed by Bicknell et
al. (1997). It is also possible that both mechanisms play a role. 

The optical identifications in all the forementioned samples are about
evenly divided into galaxies, typically in the redshift range between
0.1 and 1, and quasars, generally at higher redshift, between about
0.8 and 3.5. This implies that quasars have on average a higher
intrinsic turnover frequency.

It has been statistically found that a fraction of CSS quasars has
apparent sub-galactic size due to projection effects (Fanti et 
al. 1990), and further, the pc-scale radio morphology seems to
distinguish between galaxies (compact doubles or Compact Symmetric
Objects, CSOs) dominated by
mini-lobe emission, and quasars, where most of the radio emission
comes from unresolved cores and pc-scale jets (e.g. Stanghellini et
al. 1997). There are a few exceptions to this picture, and those
particularly interesting are the lobe dominated quasars. In general
the radio cores of the GPS quasars are less prominent than in flat
spectrum quasars (Stanghellini et al. 2001). All this is
consistent with Unified Scheme predictions.

\begin{table*}
\label{sources}
\caption{HFP sources observed at the TNG. Columns list the Name (1),
the radio position (2,3 Right Ascension and Declinations respectively,
J2000.0), the total exposure time in seconds in the $B$ (4), $V$ (5),
$R$ (6) and $I$ (7) filters respectively.} 
\begin{tabular}{cllrrrr}
\hline
\hline
Source Name & ~~~~~RA &~~~~DEC & $B$~~ & $V$~~ & $R$~~ & $I$~~  \\
   (1)      &~~~~~~(2)&~~~~~(3)& (4)~ & (5)~ & (6)~ & (7)~  \\
            &\,h\, m\,\, s &$~^\circ\quad\!^\prime\quad\!^{\prime\prime}$ &s~~&s~~&s~~ \\
\hline
\hline
J0003+2129& 00\,03\,19.3497~&21\,29\,44.436&  -~~ & 1800 & 1890 & -~~~\\
J0037+0808& 00\,37\,32.15~~~&08\,08\,12.6  &  -~~ & 1200 & 1200 & -~~~\\
J0329+3510& 03\,29\,15.3553~&35\,10\,05.940& 180  &  180 &  180 &180 \\
J0357+2319& 03\,57\,21.6088~&23\,19\,53.859& 300  &  300 &  900 & -~~~\\
J0428+3259& 04\,28\,05.8126~&32\,59\,51.975&  -~~ &  600 &  900 & -~~~\\
J1811+1704& 18\,11\,43.1843~&17\,04\,57.279&  -~~ & 1200 & 1200 & -~~~\\
J1855+3742& 18\,55\,27.7057~&37\,42\,56.986&  -~~ & 1200 & 1200 & -~~~\\
J2021+0515& 20\,21\,35.2802~&05\,15\,04.770&  -~~ &  600 &  900 & -~~~\\
J2024+1718& 20\,24\,56.5638~&17\,18\,13.208&  60  &   30 &   60 & -~~~\\
J2114+2832& 21\,14\,58.3340~&28\,32\,57.206& 600  &  600 &  600 &600 \\
J2203+1007& 22\,03\,30.9534~&10\,07\,42.584&  -~~ & 1800 & 1950 & -~~~\\
J2207+1652& 22\,07\,52.8674~&16\,52\,17.837& 180  &  180 &  180 & -~~~\\
J2212+2355& 22\,12\,05.9680~&23\,55\,40.591& 180  &  180 &  180 & -~~~\\
\hline
\hline
\end{tabular}
\end{table*}

The currently most accepted model (known as the ``youth model'') to
interpret these intrinsically small radio sources postulates that they
are indeed young objects still developing/deploying their radio
emission within the host galaxy, and their fate is to grow up to
become extended sources (hundreds of kpc in size), stage where they
spend most of their life (Fanti et al. 1995; Readhead et al. 1996),
progressively decreasing their radio power by about one order of
magnitude. A self-similar evolutionary model for CSS/GPS radio
galaxies has been proposed by Snellen et al. (2000) using a slightly
different parametrization.
\hfill\break  
The alternative model (the ``frustration model'') suggests that the
radio emitting regions would never expand due to a particularly dense
medium impeding the expansion of the radio lobes (e.g. van Breugel
1984; Baum et al. 1990). \hfill\break                                      

So far, various observations across the whole electromagnetic spectrum
failed to reveal the presence of such particularly dense medium,
suggesting, otherwise, that the host galaxies of these compact sources
share the same properties of the hosts of extended radio sources
(e.g. Fanti et al. 2000). \hfill\break
Further, the observation of proper motions of the outer edges
(hot-spots) in Compact Symmetric Objects (CSOs), a subclass of CSS/GPS
radio sources, discovered by Owsianik, Conway \& Polatidis (1998)
definitely support the youth model. Nowadays, proper motions with
apparent velocities of 0.2-0.4$c$ have been discovered in about ten
sources (Fanti 2000). \hfill\break
The radiative ages estimated for these small sources are consistent
with the hypothesis that they are young (Murgia et al. 1999).

There is a correlation between the turnover frequency and the projected
linear size (i.e. age), and the youngest (i.e. the smallest) objects
have the highest turnover frequencies (e.g. O'Dea 1998); the peak in
the radio spectrum moves at lower frequencies as the source expands
and the energy density decreases. Searching for newly born radio
sources means to look for radio spectra peaking at a few GHz or higher
frequencies. In principle, the higher the turnover frequency the
younger the object.

\section{The bright HFP sample}

To search for very young radio sources we are investigating a sample
of 55 bright High Frequency Peakers (HFPs) selected by comparing the
NRAO VLA Sky Survey, NVSS, (Condon et al. 1998) and the Green Bank,
87GB, (Gregory et al. 1996) radio catalogues at 1.4 and 4.9 GHz,
respectively. Details on the selection procedure and on the final HFP
source list can be found in Dallacasa et al. (2000).  

We used the NASA/IPAC Extragalactic Database (NED) to investigate the
current identification and redshift status of the the 55 bright HFP
sources. This search provided the following information: 29 already
have both optical identification and redshift. Of these 24 are
quasars, 1 is classified as ``stellar'' with an uncertain redshift,
and therefore likely to be also a quasar, 3 are galaxies (one of them
is a broad line galaxy) and the remaining object is a BL Lac. 
Further, either in the literature or on the red digitized Palomar
Observatory Sky Survey POSS-I images 5 have
``stellar'' appearance, 1 is a BL Lac without redshift and 4 more are
likely to be galaxies. Finally 16 are Empty Fields (EF); the summary
of the previous status of the optical identification for the whole
sample is given in Dallacasa et al. (2000). We note that a few sources
appearing as empty fields in the POSS-I plates have a weak counterpart
in the POSS-II. 

\begin{table*}
\label{optical}
\caption{Optical positions and magnitudes from TNG observations: the
various columns report the name (1), Right Ascension (2) and
Declination (3) in J2000.0, differences between the optical and radio
positions in arcsecond, $B$ (6), $V$ (7), $R$ (8) and $I$ (9) magnitudes,
and $V - R$ (10), optical morphology (11), for resolved (R) or
unresolved (U) objects; when a '?' is present, the classification is
uncertain}
\begin{tabular}{cllrrcccccc}
\hline
\hline
Source Name & ~~~RA&~~DEC
&$\Delta_{RA}$&$\Delta_{DEC}$&$B$&$V$&$R$&$I$&$V - R$& Opt. Morph.\\ 
   (1)      & ~~~(2) &~~~~(3) &(4)~~&(5)~\,~& (6) & (7) &  (8) & (9) & (10) & (11)  \\
            &\,h\, m\,\, s
&$~^\circ\quad\!^\prime\quad\!^{\prime\prime}$ &
$^{\prime\prime}~~$&$^{\prime\prime}~~$&&&&&& \\
\hline
\hline
J0003+2129&00\,03\,19.360&21\,29\,43.83& 0.15&-0.61&  -  & 20.58& 19.65 &  -  & 0.93 & R \\
J0037+0808&unidentified  &unidentified & -~~~& -~~~&  - &$>\!24.5$\,&$>\!24.5$\,&  -  & ....&....\\
J0329+3510&03\,29\,15.313&35\,10\,05.91& 0.52&-0.03&16.26& 16.06& 16.06 &15.74& 0.00 & U \\ 
J0357+2319&03\,57\,21.635&23\,19\,54.28& 0.36& 0.42&18.82& 18.23& 17.76 &  -  & 0.47 & U \\
J0428+3259&04\,28\,05.867&32\,59\,52.35& 0.68& 0.38&  -  & 20.23& 19.25 &  -  & 0.98 & R \\
J1811+1704&18\,11\,43.176&17\,04\,57.31&-0.12& 0.03&  -  & 20.52& 20.03 &  -  & 0.49 & U \\
J1855+3742&18\,55\,27.638&37\,42\,56.64&-0.80&-0.35&  -  & 21.11& 20.70 &  -  & 0.41 & ~~R?\\
J2021+0515&20\,21\,35.297&05\,15\,04.85& 0.25& 0.08&  -  & 20.70& 21.08 &  -  &-0.38 & U \\
J2024+1718&20\,24\,56.571&17\,18\,13.52& 0.10& 0.32&18.06& 17.64& 17.55 &  -  & 0.09 & U \\
J2114+2832&21\,14\,58.348&28\,32\,57.26& 0.18& 0.05&18.70& 18.38& 18.35 &17.93& 0.03 & U \\
J2203+1007&22\,03\,30.932&10\,07\,42.81& 0.04&-0.16&  -  & 23.53& 22.28 &  -  & 1.25 & R \\
J2207+1652&22\,07\,52.843&16\,52\,17.75&-0.35&-0.09&20.25& 20.27& 19.91 &  -  & 0.36 & U \\
J2212+2355&22\,12\,05.978&23\,55\,40.79& 0.14& 0.20&18.21& 19.17& 18.70 &  -  & 0.47 & U \\
\hline
\hline
\end{tabular}
\end{table*}

\section{The optical data}
On September 1$^{st}$ and 2$^{nd}$ 2000, we observed 13 HFP sources
from the bright sample with the Device Optimized for the LOw
RESolution (DOLORES) on the 3.6m Telescopio Nazionale Galileo (TNG)
located in La Palma (Canary Islands, Spain). The instrument was used
in imaging mode, operating a Loral thinned and back-illuminated
2048x2048 CCD, with a scale of 0.275 arcsec/px, yielding a field of
view of about 9.4 x 9.4 arcmin.   

All the sources listed in Table 1 were observed in R and V
filters. The exposure times were initially estimated on the basis of
the appearance of the field in the digitized POSS-I images. The
sources with a clear counterpart in such fields were also observed
with the B filter, while for two of them also the I frame was
acquired. Table 1 reports the total exposure time per filter. R
and V filters are in the Cousins system.

Five out of 13  sources (J0329+2129, J2024+1718, J2114+2832,
J2207+1652 and J2212+2355) have a counterpart in the digitized
POSS-I and/or POSS-II images, generally of stellar appearance,
although a proper classification is not possible, given that they are
rather weak.  

The radio positions in Table 1 are taken from the Jodrell Bank -- VLA
Astrometric Survey (JVAS) (Patnaik et al. 1992, Browne et al. 1998,
Wilkinson et al. 1998) and are very accurate.  
J0037+0808 is not included in the JVAS catalogue, and in Table 1 we
report the position found in the NVSS catalogue with about 1 arcsec
accuracy.  

The data reduction followed standard procedures (de-biassing, flat
fielding) in each filter within the National Optical Astronomical
Observatory (NOAO) IRAF package. The photometry was calibrated by
using the standard stars (Landolt, 1992) in two fields (SAO110-503 and
PG 2331+055A), observed each night at different air masses. Both nights
had about the same good photometric quality, and the photometry has
an accuracy of about 0.05 magnitude in the R, B and V filters, 0.1 
magnitude in the I filter.  \hfill\break
All the magnitudes were corrected for galactic extinction according
to Schlegel et al. (1998), while no correction was made for the
atmospheric contribution, which was generally within the accuracy of
the photometry mentioned above.

The seeing turned out to be quite stable in both nights ranging from
1.0 to 1.3 arcsec.
\begin{figure}
\includegraphics{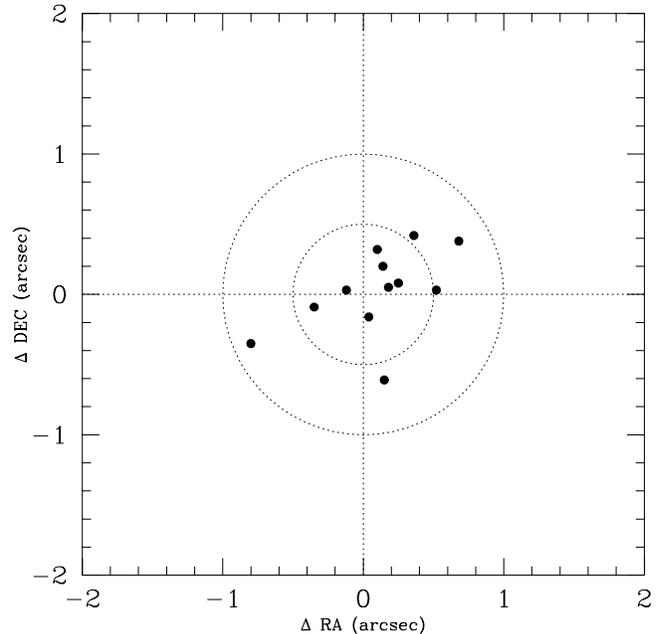} 
\vspace{8.0cm}
\caption{ 
Difference between RA and DEC derived from the optical and radio
optical positions. The two dotted circles have radius of 0.5 and 
1.0 arcsec.} 
\end{figure}

\begin{figure*}
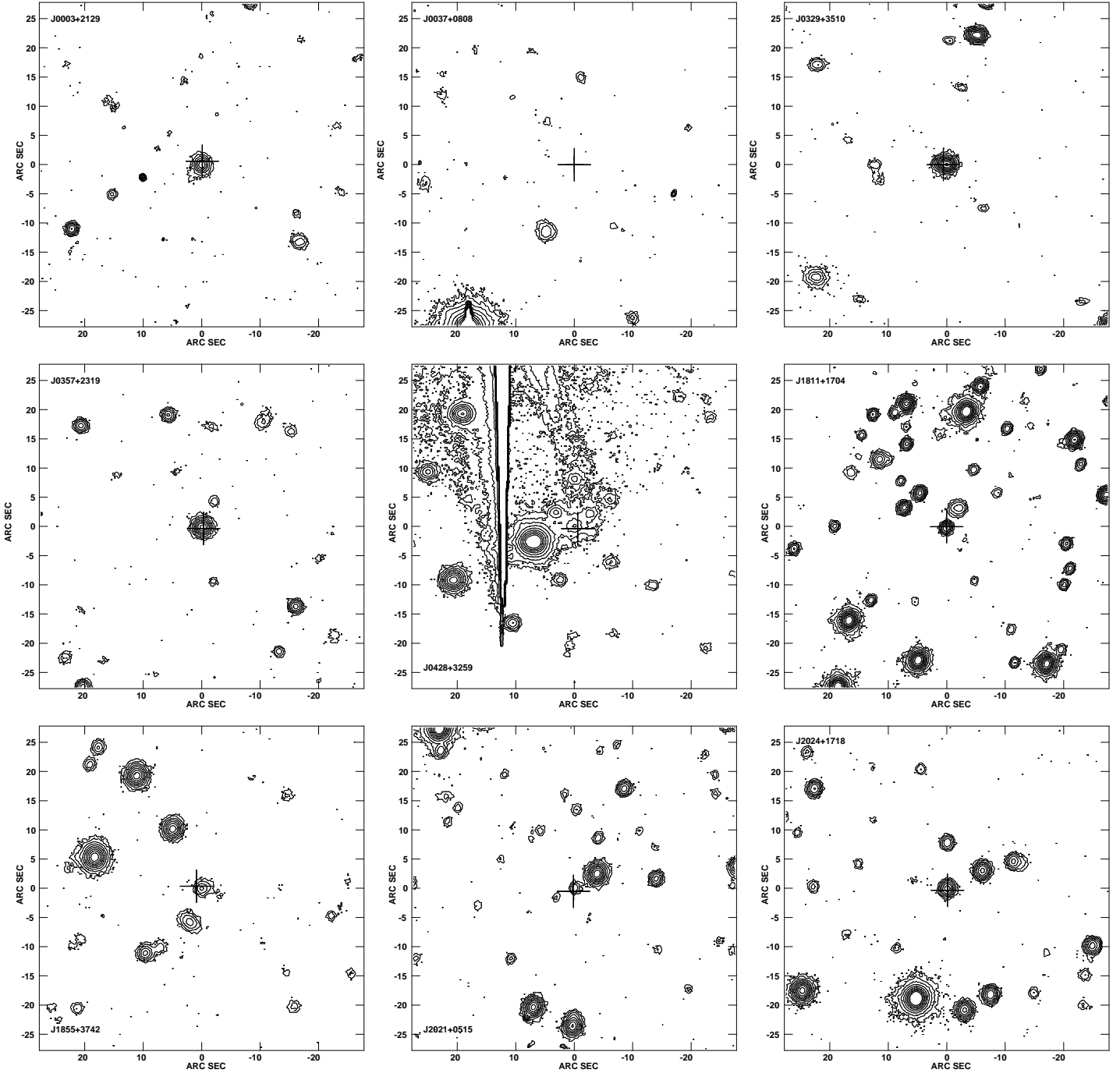

\includegraphics{H3171f2a.PS} 
\includegraphics{H3171f2b.PS} 
\includegraphics{H3171f2c.PS} 
\includegraphics{H3171f2d.PS} 
\includegraphics{H3171f2e.PS} 
\includegraphics{H3171f2f.PS} 
\includegraphics{H3171f2g.PS} 
\includegraphics{H3171f2h.PS} 
\includegraphics{H3171f2i.PS} 
\vspace{19.0cm}
\caption{
R-Band images of HFPs. Only the inner part of the CCD frame is
presented. The HFP host is always at the center of the image, and the
radio position is marked by a cross. North is up, East is to the left.} 
\end{figure*}

\section{Results}
\subsection{ Optical identification}
The optical astrometry was derived from a number of GSC stars present
in each field around the HFP source. The errors in this procedure have
been estimated to be smaller than the uncertainty between the radio
and optical reference frames, and therefore neglected.
The optical identification was successful for all sources except
J0037+0808. 

A fit to the position of each host object was carried out in each
filter. In order to have a homogeneous criterion, we decided to
consider ``optical position'' the average of the RA and DEC obtained
for the R and V images. The difference between these positions is
generally within 0.1 arcsec. Two sources have slightly larger
differences:
\begin{itemize}
\item{} J2021+0515: the host lies near a much brighter source (about
$5''$, with $R$=17.46) and the $\sim 0.3''$ difference between the
position in the two filters can be explained by the contamination from
this source
\item{} J2203+1007: it is the weakest object identified in the present
work, and its emission is clearly extended, without an outstanding
peak, and partially contaminated by a nearby extended object, possibly
a companion of the host of the HFP source.
\end{itemize}

We also measured the FWHM on the red images for all the 12 HFP
hosts. The resulting angular sizes were then compared to the FWHM of
stars in nearby regions in the same frame in order to estimate the
extension of the HFP host and complement the information derived from
the color index ($V-R$). The classification was into two main
classes, i.e. resolved ('R') or unresolved ('U') as reported in column
11 in Table 2. A question mark following the letter means that
classification is uncertain.

\setcounter{figure}{1}

\begin{figure*}
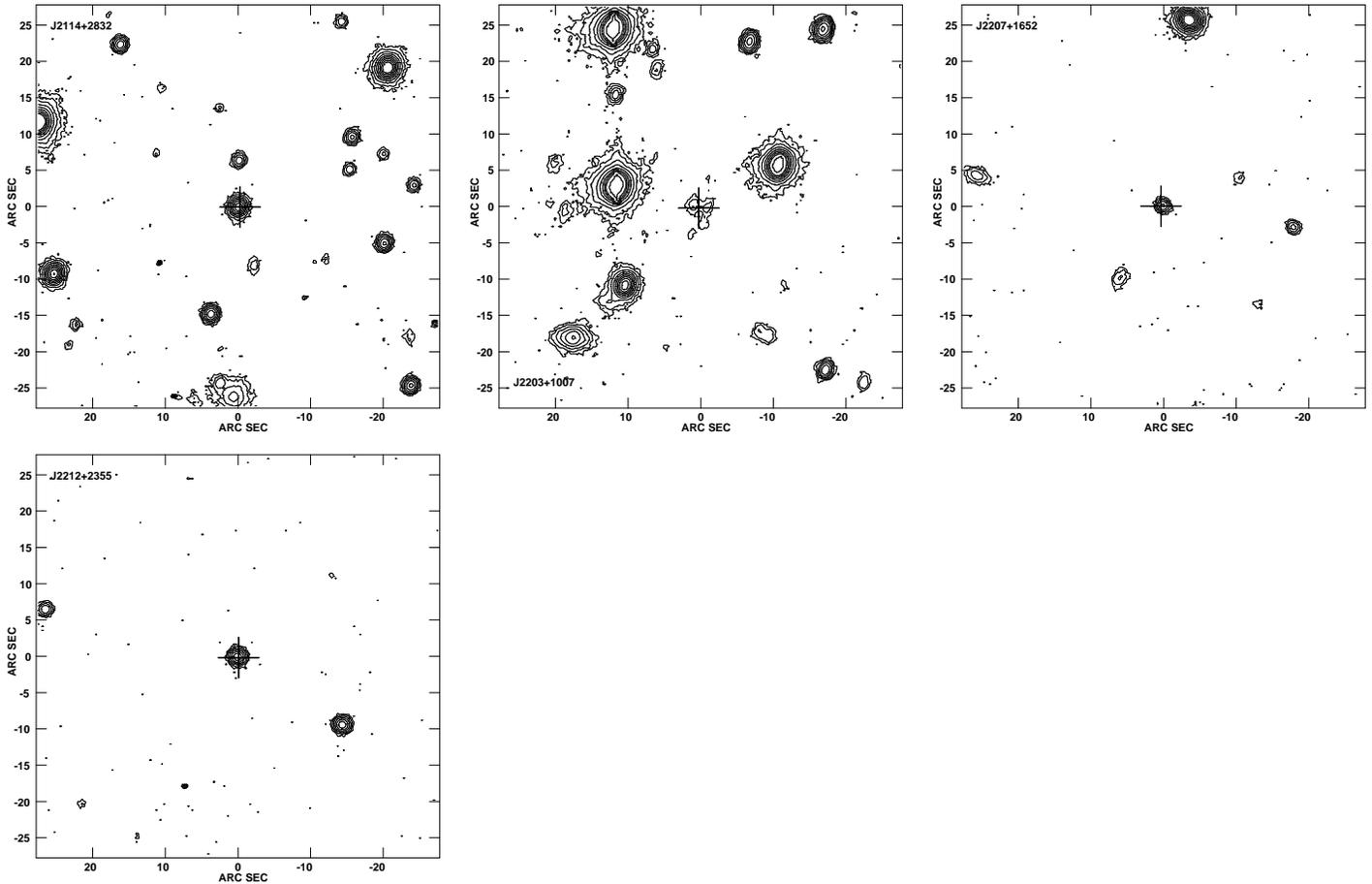

\includegraphics{H3171f2j.PS} 
\includegraphics{H3171f2k.PS} 
\includegraphics{H3171f2l.PS} 
\includegraphics{H3171f2m.PS} 
\vspace{12.6cm}
\caption{\it (continued)}
\end{figure*}

\subsection{Notes on individual sources}
For a few sources we also comment on the POSS-II (R and B) images,
characterized by some increase in sensitivity with respect to the
POSS-I (R) images, when this information is relevant to outline the
optical properties of the HFP host.

{\sl J0003+2129}\hfill\break
The radio source is coincident with the brightest optical object
within $25''$ from the radio position. We measured $R=19.65$, and
$V-R=0.93$. The lowest countour in Fig. 2 supports the hypothesis
that this object is likely a galaxy. Indeed, also the FWHM measured
for the HFP host is about 14\% broader than that determined for the
stars. \hfill\break
If we consider the Hubble diagram (Snellen et al. 1996), we can
estimate a redshift of 0.4. No optical counterpart is visible in the
POSS-I image, while very weak emission is present in both R and B
POSS-II images. It is likely that the optical emission of this object
is variable, and has increased since the POSS-I and POSS-II
observations. 

{\sl J0037+0808}\hfill\break
No significant emission has been detected in our images. No
counterpart is visible in either POSS-I or POSS-II images. 
From the present observation, the upper limit to the R magnitude is
$m_{R} > 24.5$.  By assuming that the host is a galaxy and by using
the forementioned relation we can set up a lower limit to the redshift
of this object, likely to be at $z>1.8$. 

{\sl J0329+3510}\hfill\break
POSS-I and POSS-II images show a counterpart with stellar appearance.
The TNG data confirm that the radio source is identified with a bright
stellar object with a rather flat optical spectrum and with FWHM
similar to those of the stars in the field. The R band image in Fig. 2
is clearly affected by tracking problems, since all the point sources
appear elongated approximately in E-W direction.
There is a number of extended objects in the field in
a narrow magnitude range about 4.5 mag weaker than the HFP host.

{\sl J0357+2319}\hfill\break
The optical host is definitely a variable source. No counterpart can
be found in the POSS-I images, while a weak object is visible in the
POSS-II images just at the limit allowed by the sensitivity of the
survey. Our observations revealed a relatively bright, unresolved
object with $R=17.76$. There are various extended objects in the
field  around J0357+2319, spanning a narrow magnitude range about 4.5
mag weaker than the HFP host.

{\sl J0428+3259}\hfill\break
The optical identification is with an extended object of
$R=19.25$ located very close to the radio position indicated by a
cross in Fig. 2. The field is indeed rather complex, due to the presence
of a number of nearby extended sources with magnitude similar to the
HFP host or slightly weaker. The stray pattern of a bright star about
$40''$ North of the radio source and another star about $7''$ to the
West increase the local noise in the region around the HFP
source. \hfill\break 
If we refer to the Hubble diagram in Snellen et al. (1996) we can
estimate a redshift of 0.3 for this galaxy. 

{\sl J1811+1704}\hfill\break
The optical field around the radio source is crowded by a number of
objects with similar magnitude. The proposed optical identification is
with an unresolved source sitting at about 0.1 arcsecond from the
radio position. There is an extended object 0.35 mag weaker at
about $4''$ to the North-East. The HFP host is visible in the POSS-II
red image, but it is not detected in the blue image.

{\sl J1855+3742}\hfill\break
The optical identification is with an object with $R=20.70$.
Its FWHM is a few percent broader than the average value for the stars
in this frame; in addition to this, some weak extended emission is
visible West of the radio position, and possibly connected with the
HFP host. An extended object with $R=20.29$ is located about $6''$ to
the South. A weak emission is visible in both red and blue POSS-II
images in the position of the HFP host, while no significant emission
can be revealed in the less sensitive POSS-I red image.
If we consider the HFP host a galaxy, its redshift can be estimated to
be about 0.5.

{\sl J2021+0515}\hfill\break
No counterpart is visible either in POSS-I or in POSS-II images given
the presence of a nearby stellar emission extending up to the radio
position. Our TNG images have a better resolution (the seeing is about
1.2 arcsec) and the optical host of the HFP source is a rather weak
object with $R=21.08$ whose FWHM is the same of the nearby
stars. This object is carachterized by $V - R = - 0.38$.
A number of weak extended objects about 0.5-1 mag ($R$) weaker
than the HFP host are present in this field (see Fig. 2).

{\sl J2024+1718}\hfill\break
This HFP radio source has an optical identification already, with a
stellar object, whose redshift of 1.050 is reported without reference
in the NED database. The comparison between the red and blue POSS-II
images reveals that the HFP host is the bluest of the three objects in
a line. \hfill\break
Our TNG data confirm the optical identification with an unresolved
object with a rather flat optical spectrum. 

{\sl J2114+2832}\hfill\break
A clear counterpart of stellar appearance is visible in all POSS
images. Our TNG data confirm the optical identification is with an
unresolved object with $R=18.35$ and a rather flat optical
spectrum. 

{\sl J2203+1007}\hfill\break
No counterpart is visible in any of the POSS images.
The radio source position falls within two peaks of extended
emission. We assume that the HFP host is the eastern object, being
closer. Indeed there is some additional extended emission to the
South, suggesting that there could be some interaction between
galaxies.

This is the weakest object detected in this work, and is very likely
to be a galaxy, given that $V - R = 1.25$. Again, by using the
Hubble diagram in Snellen et al. (1996) we can estimate the redshift
for this source to be about 0.9.

{\sl J2207+1652}\hfill\break
A weak counterpart is visible in all POSS images.
The optical identification is with an unresolved source with
$R=19.91$ and a rather flat optical spectrum. This source is somewhat
weaker than the counterpart visible in the POSS-I and POSS-II images.

{\sl J2212+2355}\hfill\break
The host of the radio source has a stellar profile and is much
brigther in the B filter than in V and R. Also in the digitized
POSS-II images the counterpart is significantly brighter in the blue
than in the red plates.  

\section{Summary and Future Work}

We have presented optical identifications for 12 out of 13
objects from a sample of 55 bright HFP radio sources. One of them 
has not been detected; it is likely a distant galaxy.  \\
It turns out that five sources of the 13 observed (all empty fields on
the POSS plates) can be classified as galaxies. \\
The remaining 8 objects, being unresolved in all our images and with
rather blue colors, are likely distant quasars.

Based on the results of this paper and the identifications available
in the literature only 12 out of the 46 HFPs with optical
identification are associated to galaxies (26\%),
and 34 to stellar objects (74\%).\\
Nine HFP sources do not an have optical identification yet (five
are empty fields on the POSS plates). 
Assuming they have the same proportion between extended and unresolved
morphologies found in these observations we expect that about half of
them will be galaxies.
This will raise the fraction of galaxies to about 30\%
This fraction is definitely lower than that in previous samples of
bright CSS and GPS sources (Fanti et al. 1990; O'Dea 1998;
Stanghellini et al. 1998).

Two HFP hosts, namely J0428+3259 and J2203+1007, appear to
belong to small groups, with possible interaction, although further
investigation to proof the physical connection with the other objects
is required. 

In a number of cases there are several weak galaxies with magnitudes
distributed in a narrow range (0.5-1.0) located at small angular
distances (within 30 arcsec) from the HFP host. However they are
significantly weaker than the stellar HFP host, while they have
comparable magnitudes in case of the galaxy J0428+3259. 

Optical identification of the nine sources not yet imaged and redshift
determination are needed in order to complete the classification and
to derive the rest frame parameters of all sources.

Radio imaging with (sub-)milliarcsecond resolution in the optically
thin region of the spectrum has been planned (VLBA time has been
awarded) for the majority of the sources where the morphological
information is missing or inadequate. Among these sources there are a
few known Compact Symmetric Objects (CSOs), while a few more require
confirmation. 

\begin{acknowledgements}
This research has made use of the NASA/IPAC Extragalactic Database
(NED) which is operated by the Jet Propulsion Laboratory, California
Institute of Technology, under contract with the National Aeronautics
and Space Administration. \\ IRAF is distributed by KPNO, NOAO,
operated by the AURA, Inc., for the  National Science Foundation.
\end{acknowledgements}

{} 

\end{document}